\documentclass[aps,prl,oneside,twocolumn]{revtex4-1}
\usepackage{amsmath}
\usepackage{amssymb}
\usepackage{graphicx}
\usepackage{color}

\begin{document}
\draft \title{A classical model of the upper bounds of the cascading contribution to the second hyperpolarizability}
\author{Nathan J. Dawson$^{\ddag,\dag,1}$, Benjamin R. Anderson$^{\ddag}$, Jennifer L. Schei$^{\ddag}$, and Mark G. Kuzyk$^{\ddag,2}$}
\address{$^{\ddag}$Department of Physics and Astronomy, Washington State University,
Pullman, WA 99164-2814 \\ $^{\dag}$ Currently at Department of Physics and Astronomy, Youngstown State University, Youngstown, OH 44555}
\email{$^1$dawsphys@hotmail.com, $^2$kuz@wsu.edu}
\date{\today}

\begin{abstract}We investigate whether microscopic cascading of second-order nonlinearities of two molecules in the side-by-side configuration can lead to a third-order molecular nonlinear-optical response that exceeds the fundamental limit. We find that for large values of the second hyperpolarizability, the side-by-side configuration has a cascading contribution that lowers the direct contribution. However, we do find that there is a cascading contribution to the second hyperpolarizability when there is no direct contribution. Thus, while cascading can never lead to a larger nonlinear-optical response than for a single molecule with the same number of electrons, it may provide design flexibility in making large third-order susceptibility materials when the molecular second hyperpolarizability vanishes.
\end{abstract}


\maketitle

\section{Introduction}

Cascading is a process in which the interactions of light beams generated in lower-order nonlinear optical processes lead to a higher-order nonlinear-optical response.  The lowest-order cascading phenomena is an effective third-order susceptibility that results from interactions of fields produced by two second-order nonlinear susceptibilities.  The microscopic analogue is a second hyperpolarizability that results from interactions of fields produced by two molecules through their hyperpolarizabilities.

\begin{figure}
\centering\includegraphics{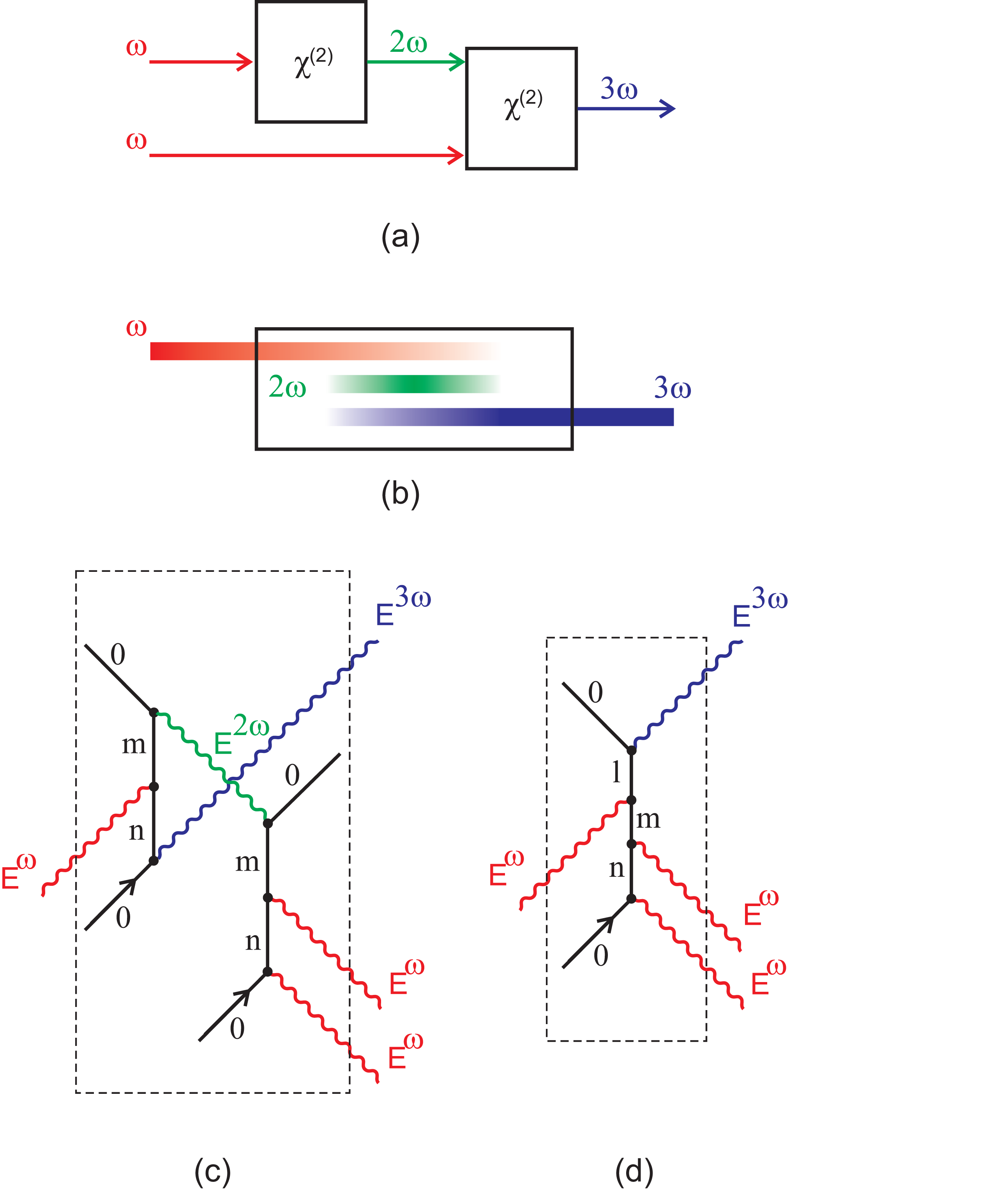}
\caption{(Color online) (a) Third harmonic generation from two second-order processes; (b) phase-matched third harmonic generation in a two step process; (c) a Feynman diagram for one configuration of microscopic cascading; and, (d) the direct third order process of third harmonic generation.}
\label{fig:cascading}
\end{figure}

Figure \ref{fig:cascading}(a) shows how third harmonic light can be generated using two separate crystals through their second-order susceptibilities.  Similarly, within a material that lacks inversion symmetry, monochromatic light of angular frequency $\omega$ is converted to second harmonic light at frequency $2 \omega$, which through parametric mixing with the fundamental leads to the generation of third harmonic light at frequency $3 \omega$.  Figure \ref{fig:cascading}(b) illustrates that, from the perspective of an observer outside the crystal, the process appears as third-harmonic generation due to a third-order nonlinear-optical susceptibility. Cascading of nonlinear susceptibilities was used early on by Coffinet and Martini to characterize coherent excitation of polaritons in Gallium Phosphide.\cite{pierr69.01}

Figure \ref{fig:cascading}(c) shows one of the Feynman diagrams for cascading between two molecules (time runs from bottom to top). The effective third-order susceptibility is calculated by summing over all topologically distinct diagrams.\cite{kuzyk85.01} To an outside observer, who is not privy to the virtual processes involved inside the dashed box, the cascading third-order susceptibility is indistinguishable from a direct third-order susceptibility, shown in Figures \ref{fig:cascading}(b) and \ref{fig:cascading}(d).  Microscopic cascading can lead to a third-order susceptibility even in an isotropic material,\cite{kuzyk85.01} and is related to the pair correlation function.\cite{andre91.01,andre92.02,baev10.01}

Yablonivitch showed that cascading contributes substantially to the third-order susceptibility in GaAs\cite{yablo72.01} while Meredith and Buchalter studied the contributions of cascading in an $\alpha$-quartz crystal using wedge fringing,\cite{mered81.07} and Meredith measured third harmonic generation of para nitroaniline (pNA) in solution as a function of concentration to determine the effects of microscopic cascading.\cite{mered83.03}  Meridith concluded that most of the observed third harmonic light for molecules with a large hyperpolarizability, such as pNA, is due to microscopic cascading, and that the hyperpolarizability needs to be taken into account if the direct contribution of the second hyperpolarizability is to be determined.  Later, Torruellas and coworkers showed that cascading was important in Alkoxy-nitro-stilbene (MONS) and Di-alkyl-amino-nitro-stilbene (DANS) side chain substituted polymers.\cite{torru91.01x}

These earlier studies focused on third harmonic generation or two beam mixing (at frequencies $\omega_1$ and $\omega_2$) that yield light at $2 \omega_1 -\omega_2$.  Since optical switching devices required materials with a large intensity-dependent refractive index, subsequent studies - such as the ones by Stegeman and coworkers \cite{stege93.01,stege96.01} - considered the role of cascading in self-action effects such as the nonlinear phase shift.  A switching device based on non-phase-matched second harmonic cascading was reported by Assanto and coworkers.\cite{assanto93.01x} Bosshard and coworkers demonstrated in kNbO$_3$ that in addition to cascading due to second harmonic, the intermediate field could be static, as one finds in the electrooptic effect and optical rectification.\cite{bossh95.01}

The earliest research on cascading focused on understanding the contributions of cascading to measurements of the third-order response of the material so that the direct response could be determined.  Also, the cascading contribution was used to estimate the second-order response, which could be compared with direct measurements of the second-order response.  Later, when it was recognized that cascading's and microscopic cascading's contributions could exceed the direct contribution, researchers exploited this fact to demonstrate practical applications that required ever-higher third-order susceptibilities.

Recently, Baev and coworkers have developed a phenomenological model for microscopic cascading with the goal of introducing new design principles for making nonlinear materials with larger third-order nonlinearity.\cite{baev10.01} Thus, it is natural to ask whether two molecules with large hyperpolarizabilities can have a larger second hyperpolarizability due to cascading effects than is possible for individual molecules. This question can be answered using the theory of fundamental limits of a quantum system,\cite{kuzyk00.02, kuzyk00.01} which is normally applied to single molecules, to determine whether or not cascading provides a loophole for breaking the single-molecule limits.

Based on the fact that the theory of fundamental limits \cite{kuzyk00.02,kuzyk00.01,kuzyk03.01,kuzyk03.02,kuzyk03.03} applies to any quantum system - independent of the details, one would anticipate that if the two-molecule system is considered as the quantum system, then the fundamental limits must still hold for cascading.  Our paper is thus separated into testing two hypotheses.  (1) We calculate the cascading contribution to the third-order nonlinearity for a two-molecule system in a side-by-side geometry along the electric field lines where the nonlinear contribution is expected to be optimal, and test the hypothesis that the fundamental limit of this system is not breachable.  (2) When the direct third-order response is small, we test the hypothesis that cascading can lead to an appreciable third-order nonlinearity.

The first hypothesis is validated.  With regards to the second hypothesis, we find that for molecules with a large second-order nonlinearity, the cascading contribution can be large, but still falls far short of the fundamental limit due to the direction of the local electric field.  Whether the constraints imposed on real molecules will lead to larger direct second hyperpolarizabilities or cascading hyperpolarizabilities is yet to be seen.  However, given that small molecules with second hyperpolarizabilities near the fundamental limit have already been observed,\cite{May05.01,May07.01} cascading may be more promising in the design of molecules with larger susceptibilities.  We conclude our work by discussing these scaling issues.

\section{Theory}
\label{sec:configurations}

In this section, we calculate the electric field due to an arbitrarily-oriented dipole at the location of a second dipole - taking into account back reaction using the method of self-consistent fields.  Then, treating a material as a collection of molecules in a lattice, we calculate the electric field at at two neighboring lattice points due to dipole pair interactions.  We focus on the classical model of cascading, which is defined by the fact that the molecular properties such as the polarizability and hyperpolarizabilities remain unchanged by local field interactions, to which cascading contributes.

The classical model is well-behaved only for the side-by-side geometry.  As we will show in the companion article,\cite{dawson11.02a} the classical model breaks down in the end-to-end configuration because changes in the optical properties of the molecules must be taken into account to offset the effects of divergence.  This requires the introduction of quantum effects, which will be treated in the companion article using perturbation theory.

Since our focus is on understanding the largest possible cascading contribution, we treat only the cases that should yield the largest response, i.e. those that approach the fundamental limit.  Our theory can be easily modified to other cases, but we do not do so here because this would not address the question of enhancement of the nonlinear-optical response at the extremes.

\subsection{Two one-dimensional molecules}
\label{sub:twomolecules}

Consider two randomly oriented molecules that are cylindrically symmetric and subject to a uniform applied electric field. We define the laboratory frame's $z$-axis to be parallel to the static electric field, where the cartesian coordinates for the laboratory frame are $x$, $y$, and $z$. We denote the body frame's coordinates as $x_{i}'$, $y_{i}'$, and $z_{i}'$, where $i$ is a reference to molecule $i$, thus taking on values $1$ or $2$. The two dipoles are identical except for their orientation. Moreover, we assume that each molecule's linear and nonlinear susceptibilities are dominated by one of the diagonal tensor components, or equivalently, that the molecules are approximately one dimensional.  This is found to be a good approximation for donor-accepter molecules such as para-Nitroaniline (pNA), stilbenes, azo dyes, etc.\cite{kuzyk98.01} -  commonly used in nonlinear optics.

We can ignore the axial Euler angle due to the uniaxial symmetry of the molecule, thus, only $\phi_i$ and $\theta_i$ need to be considered. The Euler rotation matrices about the two remaining axes are of the form,
\begin{equation}
A\left(\phi_i\right) = \begin{pmatrix} \cos \phi_i & \sin \phi_i & 0 \\-\sin \phi_i & \cos \phi_i & 0 \\0 & 0 & 1 \end{pmatrix}
\label{eq:transA}
\end{equation}
and
\begin{equation}
B\left(\theta_i\right) = \begin{pmatrix} 1 & 0 & 0 \\0 & \cos \theta_i & \sin \theta_i \\ 0 & -\sin \theta_i & \cos \theta_i \end{pmatrix} .
\label{eq:transB}
\end{equation}
The full Euler transformation matrix from the body coordinates to the laboratory frame is
\begin{equation}
D\left(\phi_i,\theta_i\right) = \begin{pmatrix} \cos \phi_i & -\sin \phi_i \cos \theta_i & \sin \phi_i \sin \theta_i \\ \sin \phi_i & \cos \phi_i \cos \theta_i & -\cos \phi_i \sin \theta_i \\ 0 & \sin \theta_i & \cos \theta_i \end{pmatrix} .
\label{eq:transfull}
\end{equation}
where $D\left(\phi_i,\theta_i\right) = A^T \left(\phi_i\right)B^T \left(\theta_i\right)$. Here the superscript $T$ denotes the transpose.

The dipole moment of the $i$th molecule in the laboratory frame, $\mathbf{p}_{i}$, can now be written as
\begin{equation}
\mathbf{p}_{i} = D\left(\phi_i,\theta_i\right) \mathbf{p}_{i}' .
\label{eq:plabframe}
\end{equation}
The electric field in the laboratory frame at position $r$ due to molecule $i$ with an induced dipole moment at position $r_i$ is
\begin{equation}
\mathbf{E}_{i} = \frac{3\left(\hat{r}-\hat{r}_{i}\right) \left[\mathbf{p}_{i}\cdot \left(\hat{r}-\hat{r}_{i}\right)\right] - \mathbf{p}_{i}}{\left|\mathbf{r} - \mathbf{r}_{i}\right|^3} ,
\label{eq:labframeE}
\end{equation}
where $\mathbf{r}_{i}$ is the vector from the origin to the $i$th molecule, and $\mathbf{r}$ is the vector that points from the origin to the electric field point.

We define the $m^{th}$-order molecular susceptibility as $k^{\left(m\right)}$, where $\alpha = k^{\left(1\right)}$, $\beta = k^{\left(2\right)}$, and $\gamma = k^{\left(3\right)}$. Then the induced dipole moment of the $i$th molecule in the laboratory frame approximated to third order is
\begin{eqnarray}
\mathbf{p}_{i} &=& \hat{f}_i \left(\phi_i,\theta_i\right) \alpha \left[\hat{f}_i\left(\phi_i,\theta_i\right)\cdot \left(\mathbf{E}_a + \mathbf{E}_{j}\right) \right] \nonumber \\
&+& \hat{f}_i\left(\phi_i,\theta_i\right) \beta \left[ \hat{f}_i \left(\phi_i,\theta_i\right) \cdot \left(\mathbf{E}_a + \mathbf{E}_{j}\right)\right]^2 \nonumber \\
&+& \hat{f}_i\left(\phi_i,\theta_i\right) \gamma \left[ \hat{f}_i \left(\phi_i,\theta_i\right) \cdot \left(\mathbf{E}_a + \mathbf{E}_{j}\right)\right]^3 ,
\label{eq:molsusthirdorder}
\end{eqnarray}
where $\mathbf{E}_a = E_a\hat{z}$ is the applied electric field and
\begin{eqnarray}
\hat{f}_i \left(\phi_i,\theta_i\right)&=& D\left(\phi_i,\theta_i\right) \hat{z}_{i}' \nonumber \\
&=& \sin\phi_i \sin\theta_i \hat{x}-\cos\phi_i\sin\theta_i \hat{y} + \cos\theta_i \hat{z} .
\label{eq:coordunitshift}
\end{eqnarray}
Equation \ref{eq:molsusthirdorder} describes the induced dipole moment of a one-dimensional molecule in a system of two arbitrarily oriented and positioned molecules in an applied electric field. Because there are only two molecules, we only have two possible combinations of $i$ and $j$ for $i\neq j$ ($i=1$, $j=2$ or $i=2$, $j=1$). Note that the molecules are identical, and therefore $k_{1}^{\left(m\right)} = k_{2}^{\left(m\right)} = k^{\left(m\right)}$.

\subsection{Cubic lattice}
\label{sub:cubiclat}

A self-consistent solution of the above problem is highly complex. Such problems can be simplified by applying geometric constraints to the five degrees of freedom due to the second molecule: two rotational and three translational degrees of freedom.

One common method used to eliminate the translational degrees of freedom is to use a lattice model,\cite{lebwo72.01,priez01.01,romano86.01} in which the molecules are confined to lattice points.  For our model, we choose the simple cubic lattice.

We choose the cubic lattice's $z$-axis to be parallel to the applied electric field. To the lowest degree of approximation, only the nearest neighbor interactions are considered. There are six nearest neighbors to each lattice point: four sites in a direction perpendicular to the applied field and two sites parallel to the applied field.  We call these two types of nearest neighbor sites side-by-side and end-to-end, respectively. The distance between any two nearest neighbors, $r$, is related to the volume, $V$, of the primary cell, according to $r = \sqrt[3]{V}$. We use the volume as a parameter to study how the nonlinear susceptibility depends on the nearest neighbor separation.

The $i$th molecule's electric field along the three orthogonal lattice directions is calculated from Equation \ref{eq:labframeE}, yielding,
\begin{eqnarray}
E_{i,x} &=& \frac{p_i}{x^3} \left(2\hat{x}\sin\phi_i \sin\theta_i + \hat{y}\cos\phi_i \sin\theta_i - \hat{z}\cos\theta_i \right) , \label{eq:elecx} \\
E_{i,y} &=& \frac{p_i}{y^3} \left(-\hat{x}\sin\phi_i \sin\theta_i - 2\hat{y}\cos\phi_i \sin\theta_i - \hat{z}\cos\theta_i \right) , \nonumber \\ & & \label{eq:elecyt}
\end{eqnarray}
and
\begin{equation}
E_{i,z} = \frac{p_i}{z^3} \left(-\hat{x}\sin\phi_i \sin\theta_i + \hat{y}\cos\phi_i \sin\theta_i + 2\hat{z}\cos\theta_i \right) ,
\label{eq:elecz}
\end{equation}
where the six nearest neighbors to the site at $(x,y,z) = (0,0,0)$ for a cubic lattice are at $(x,y,z) = (\pm r,\pm r,\pm r)$.

Since we seek to calculate the maximum molecular susceptibility due to an applied field and the induced dipole field from an adjacent molecule, we chose to study the configuration in which the largest tensor component of the electron response of each molecule is aligned with the applied electric field. For the side-by-side case, the electric field due to each neighbor is
\begin{equation}
E_\perp = \frac{-p}{r^3} ,
\label{eq:electricperp}
\end{equation}
and in the end-to-end configuration, the field is given by
\begin{equation}
E_\| = \frac{2p}{r^3}
\label{eq:electricpara} ,
\end{equation}
where $\perp$ and $\|$ denote the measurement of the induced-dipole field in the direction of the applied field and at either a position perpendicular or parallel to the direction of the applied electric field.

While in this paper, we treat the case of two stationary molecules with fixed orientations, we will use the above theory in the companion article to treat the more general case.

\subsection{Approximations and Assumptions}

In this work, we restrict ourselves to the cascading contribution from molecules that are both aligned parallel to the electric field. The alignment with the electric field should yield the largest nonlinear response for each individual molecule.  Without loss of generality, we assume that the nonlinearity is nonzero only along the long axis of the molecule.  In this case, the largest cascading contribution, by either contributing or negating the direct second hyperpolarizability, corresponds to a configuration where the induced dipole moment of one molecule is along the field line due to the other induced dipole. As we find later in this paper, side-by-side cascading along the electric field lines suppresses the second hyperpolarizability and thus gives the lower bound of the cascading contribution.

Under the condition that each dipole axis is aligned along the electric field of the other dipole, there are two positions in the gas-lattice model that yield the largest response to the applied electric field: the side-by-side and end-to-end geometries. These cases will give the range of the cascading contribution.\cite{dawson11.02a} Here we focus on the side-by-side case.

We argue that the best cascading response is due to two identical dipoles. Moreover, we utilize the dipole approximation, where higher order multipole moments are ignored. Consider that a fixed amount of charge is distributed between the two dipoles so that $q_1 + q_2 = 1$.  Since the cascading effect is proportional to the product of the dipole moments, this yields $p_1 p_2 = p \left(1-p\right)$, which is optimized when the two dipoles are of equal magnitude.

When the wavelength of the light is long compared with the size of each molecule, the electric field will be approximately uniform.  Since the dipole field of a molecule is small, just a few molecular lengths away, then for typical molecules whose sizes are on the order of 1-10nm, the optical field will appear uniform over all molecular separations that will yield cascading.  Figures \ref{aligned_molecules} and \ref{antialigned_molecules} summarize the side-by-side configuration that we use in our calculations, which we argue encompasses the full domain of cascading.

It is important to stress that our focus is on microscopic cascading.  As such, the cascading process is confined to the near field, with little cascading resulting in the far field.\footnote{In contrast, macroscopic cascading results from ordered materials where the cascading fields propagate within the material.} Thus, the cascading fields can be calculated in the near zone limit where $\omega \rightarrow 0$.

\subsection{Self-consistent field calculation of interacting dipoles}
\label{sub:selfconsfieldcalc}

\begin{figure}
\centering
\includegraphics{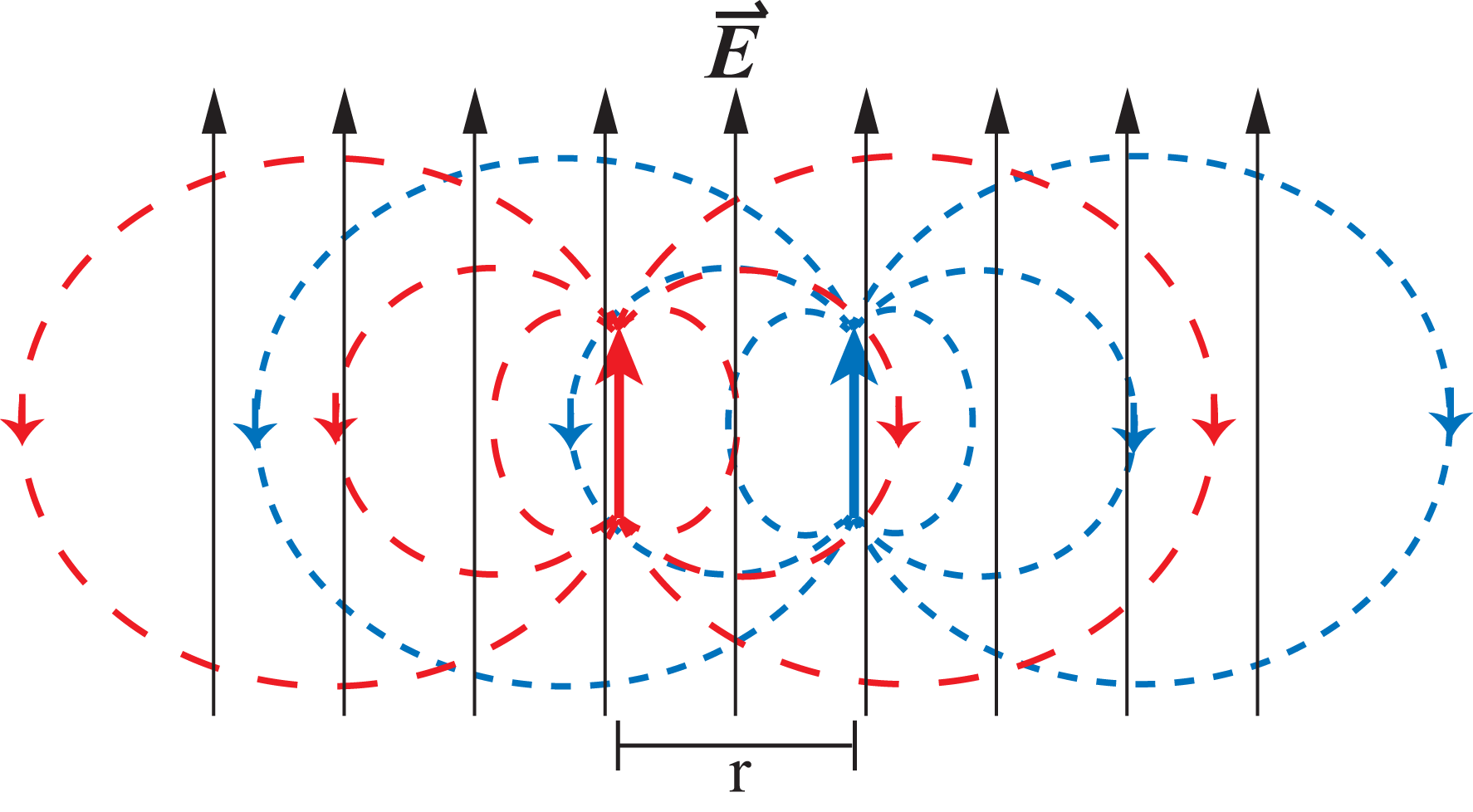}
\caption{(Color online) Two molecules are aligned in a static electric field, $\vec{E}$, and are separated by a distance $r$.}
\label{aligned_molecules}
\end{figure}
We apply a self-consistent field formulation to determine the induced dipole moment of two interacting dipoles. From Figure \ref{aligned_molecules}, it is clear that the polarization of one molecule is due to the applied electric field and the dipole field from the other molecule. The dipole field at the second molecule is given by Equation \ref{eq:electricperp}. In the zero-frequency limit, the polarizations of each molecule, to third order, are:
\begin{eqnarray}
  p_1 &=& \alpha \left({E_a}-\frac{p_2}{r^3}\right) + \beta\left({E_a}-\frac{p_2}{r^3}\right)^2 \nonumber \\
  &+& \gamma\left({E_a}-\frac{p_2}{r^3}\right)^3 , \label{eq:aligned1}
\\
  p_2 &=& \alpha \left({E_a}-\frac{p_1}{r^3}\right) + \beta\left({E_a}-\frac{p_1}{r^3}\right)^2 \nonumber \\
  &+& \gamma\left({E_a}-\frac{p_1}{r^3}\right)^3 . \label{eq:aligned2}
\end{eqnarray}
Equations \ref{eq:aligned1} and \ref{eq:aligned2} can be used to solve for $p_1$ and $p_2$ in terms of $r$ and $E_a$. Note that this self consistent approach takes into account all possible interactions between the fields.

Defining the dipole moment of the two-particle system as $p=p_1+p_2$, and recalling that the one-dimensional $n$th order molecular susceptibilities is defined as
\begin{equation}
k^{\left(n\right)} = \left. \frac{1}{n!} \frac{d^np}{dE_{a}^n} \right|_{E_a=0}, \label{eq:kn}
\end{equation}
where $k^{\left(1\right)} = \alpha$, $k^{\left(2\right)} = \beta$, and $k^{\left(3\right)} = \gamma$, we get
\begin{eqnarray}
\alpha_\mathrm{eff}^{+} &=& 2r^{3}\frac{\alpha}{r^3+\alpha} , \label{eq:aeff} \\
\beta_\mathrm{eff}^{+} &=& 2r^{9}\frac{\beta}{\left(r^3 + \alpha\right)^3} , \label{eq:beff} \\
\gamma_\mathrm{eff}^{+} &=& 2r^{12}\frac{\left(r^3+\alpha\right)\gamma - 2 \beta^2}{\left(r^3 + \alpha\right)^5}  \nonumber \\ &=& 2 \gamma \frac {r^{12}} {\left(r^3+\alpha \right)^4}  - 4 \beta^2 \frac {r^{12}} {\left(r^3+\alpha \right)^5}, \label{eq:geff}
\end{eqnarray}
where the superscript ``+'' denotes the case where both molecules are aligned with the electric field. It can be shown that $k^{-\left(n\right)} = k^{+\left(n\right)}$, where the superscript ``-'' denotes the case in which the molecules are both anti-aligned with the electric field.  We emphasize that this self-consistent approach automatically takes into account the local field at each molecular site (the factor multiplied by $2 \gamma$ in Equation \ref{eq:geff}) as well as cascading, the second term in Equation \ref{eq:geff}. The resultant cascading term is a factor of $4$ larger than that given by Baev \textit{et al}.\cite{baev10.01}

\begin{figure}
\centering
\includegraphics{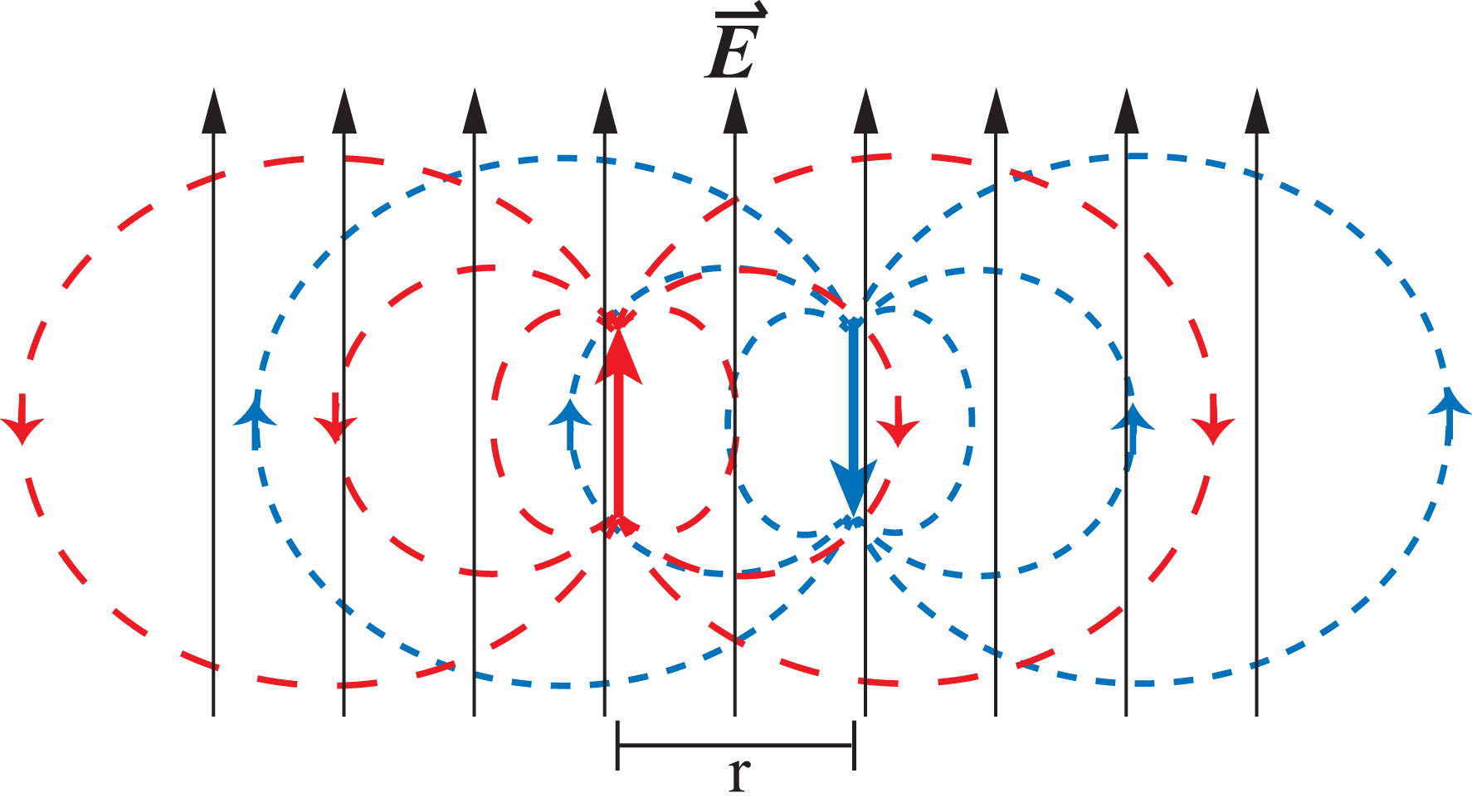}
\caption{(Color online) Two molecules in a static electric field, $\vec{E}$, are separated by a distance $r$ and are in the counter-aligned geometry.}
\label{antialigned_molecules}
\end{figure}
When the molecules are antiparallel to each other, as shown in Figure \ref{antialigned_molecules}, the polarization of each molecule is
\begin{eqnarray}
p_2 &=& \alpha \left(E_a-\frac{p_1}{r^3}\right) + \beta\left(E_a-\frac{p_1}{r^3}\right)^2 \nonumber \\
&+& \gamma\left(E_a-\frac{p_1}{r^3}\right)^3 , \label{eq:alignedanti1}
\\
-p_1 &=& \alpha\left(E_a-\frac{p_2}{r^3}\right) + \beta\left(E_a-\frac{p_2}{r^3}\right)^2 \nonumber \\ &+& \gamma\left(E_a-\frac{p_2}{r^3}\right)^3 . \label{eq:alignedanti2}
\end{eqnarray}
Substituting Equation \ref{eq:alignedanti1} into Equation \ref{eq:alignedanti2} yields
\begin{equation}
p = 0.
\label{eq:casczero}
\end{equation}
Thus, $\alpha_\mathrm{eff}^\mathrm{\pm} = 0$, $\beta_\mathrm{eff}^\mathrm{\pm} = 0$, and $\gamma_\mathrm{eff}^\mathrm{\pm}= 0$. Here, the superscript ``$\pm$'' denotes the antiparallel configuration of the two molecules. Since we are interested in studying the largest possible cascading response in the side-by-side geometry, we will only treat the parallel case.

The first term of Equation \ref{eq:geff} describes the dressed second hyperpolarizability and the second term describes the cascading contribution.  All that remains is the determination of $\alpha$, $\beta$ and $\gamma$, from which the cascading contribution can be calculated.

\subsection{The three-level ansatz}
\label{sec:3levelansatz}

Our approach is to calculate the zero-frequency nonlinearities using the sum-over states expressions, then apply the sum rules to reduce the number of parameters that are required to model the response.  The results can then be compared with the fundamental limits of the off-resonant nonlinear-optical response.  Since a rigorous theory of the on-resonance fundamental limits are also well-known,\cite{kuzyk06.03} we could apply the same approach to resonant cascading.  However, the problem is too complex to treat here.  We argue that since the off-resonant response is usually of interest in many applications, we limit our calculations to the zero-frequency limit.

The off-resonant polarizability, hyperpolarizability, and second hyperpolarizability are given by\cite{orr71.01}
\begin{eqnarray}
	\alpha &=& 2 e^2\displaystyle\sum_{n}^\infty \hspace{-.08cm}\left.\right. ^\prime \frac{x_{0n}x_{n0}}{E_{n0}} ,
\label{eq:alphasum} \\
\beta &=& 3 e^3\displaystyle\sum_{n,m}^\infty \hspace{-.08cm}\left.\right. ^\prime \frac{x_{0n} \bar{x}_{nm}x_{m0}}
{E_{n0}E_{m0}}  ,\label{eq:betasum} \\
\gamma &=& 4 e^4\displaystyle\sum_{n,m,l}^\infty \hspace{-.16cm}\left.\right. ^\prime \frac{x_{0n}\bar{x}_{nm}
\bar{x}_{ml}x_{l0}}{E_{n0}E_{m0}E_{l0}} \nonumber \\
&-& 4 e^4 \displaystyle\sum_{n,m}^\infty \hspace{-.08cm}\left.\right. ^\prime
\frac{x_{0n}x_{n0}x_{0m}x_{m0}}{E_{n0}^2E_{m0}} ,
\label{eq:gammasum}
\end{eqnarray}
respectively, which as we describe later, can be written in dipole-free form.\cite{kuzyk05.02,kuzyk09.01,perez01.08} Here, $E_{i0}=E_i-E_0$ is the energy difference between the $i^{\mathrm{th}}$ excited state and the ground state, $x_{ij}$ is the transition moment between state $i$ and state $j$, and $e$ is the electron charge. The primed sums in Equations \ref{eq:alphasum}-\ref{eq:gammasum} exclude the ground state from the summation.

The three-level ansatz states that when the nonlinear-optical response of a quantum system is at the fundamental limit, only three states contribute.\cite{kuzyk09.01}  Since we are interested in studying the limit when the direct and cascading contributions are large, we represent all of the susceptibilities using three states, i.e. the ground state and first two excited states.  In our calculations, we assume that the three-level model is a good approximation for all cases.

For a three-state model, we can explicitly write the polarizability from Equation \ref{eq:alphasum} as
\begin{equation}
	 \alpha_{3L}=2e^2\left(\frac{|x_{10}|^2}{E_{10}}+\frac{|x_{20}|^2}{E_{20}}\right) .
\label{eq:alphaexplicit}
\end{equation}
The Thomas-Kuhn sum rules relate $|x_{10}|^2$ and $|x_{20}|^2$, and thus can be used to simplify Equation \ref{eq:alphaexplicit}.  For a three-state system, the ground state sum rule takes the form
\begin{equation}
	E_{10}|x_{10}|^2+E_{20}|x_{20}|^2 = \frac{N\hbar^2}{2 m} ,
\label{eq:x20x10relation}
\end{equation}
where $\hbar$ is Planck's constant, $m$ is the mass of the electron, and $N$ is the number of relevant electrons.\cite{kuzyk01.01} It is straightforward to show from the sum rules that the maximum value of $x_{10}$ is given by \cite{kuzyk10.01}
\begin{equation}
	|x_{10}^{\mathrm{max}}|^2 = \frac{N\hbar^2}{2 m E_{10}} .
\label{eq:x10max}
\end{equation}
It is useful to define the dimensionless parameters
\begin{eqnarray}
	E &=& \frac{E_{10}}{E_{20}} ,\label{eq:Edefine}\\
	X &=& \frac{\left|x_{10}\right|}{\left|x_{10}^{\mathrm{max}}\right|} .
\label{eq:Xdefine}
\end{eqnarray}
Note that $0 \leq E \leq 1$ and $0 \leq X \leq 1$.

Using Equations \ref{eq:x20x10relation}, \ref{eq:x10max} and \ref{eq:Xdefine}, Equation \ref{eq:x20x10relation} can be expressed as
\begin{equation}
	|x_{20}|^2 = E |x_{10}^{\mathrm{max}}|^2\left(1-X^2\right) .
\label{eq:finalx10x20}
\end{equation}
Substituting Equation \ref{eq:finalx10x20} into Equation \ref{eq:alphaexplicit} gives
\begin{equation}
\alpha_{3L} = 2 e^2 \frac{|x_{10}^{\mathrm{max}}|^2}{E_{10}}\left[X^2+E^2 \left(1-X^2\right)\right] ,
\label{eq:alpha3L}
\end{equation}
and substituting Equation \ref{eq:x10max} into Equation \ref{eq:alpha3L} gives
\begin{equation}
\alpha_{3L} = \frac{e^2 \hbar^2 N}{m E_{10}^2}\left[X^2+E^2 \left(1-X^2\right)\right] .
\label{eq:alpha3N}
\end{equation}

The three-state polarizability, $\alpha_{3L}$, is maximum when the term in the brackets from Equation \ref{eq:alpha3N} equals unity.  Thus, the fundamental limit of the polarizability, $\alpha^{\mathrm{max}}$ is given by,
\begin{equation}
\alpha^{\mathrm{max}} = \frac{e^2 \hbar^2 N}{m E_{10}^2} .
\label{eq:alphaMAX}
\end{equation}
Note that Equation \ref{eq:alphaMAX} is general.  When more than two excited states are present, the oscillator strength is spread over more states, and the polarizability is smaller.  Thus, the polarizability can never exceed the fundamental limit.  This is the motivation for the three-level ansatz, which is also applied to higher-order nonlinearities.

\begin{figure}
\centering\includegraphics{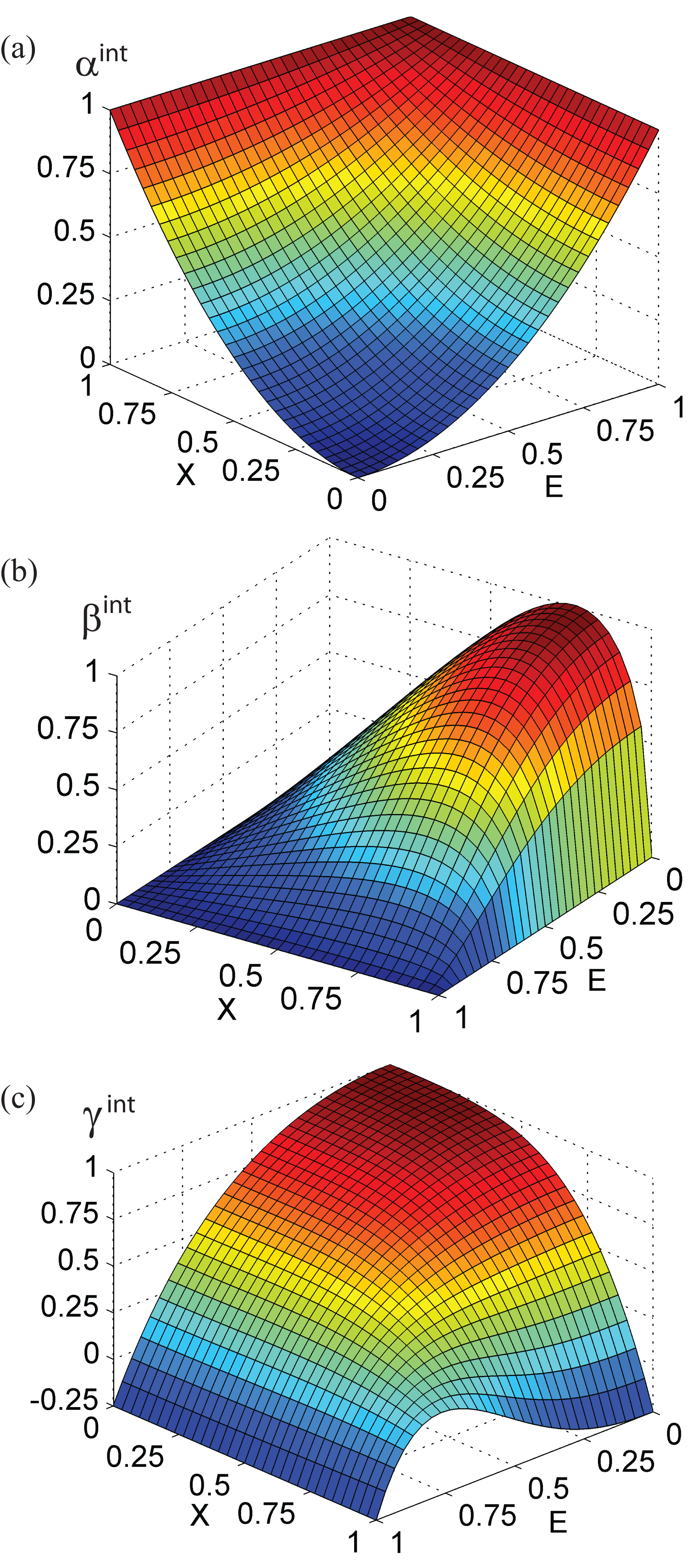}
\caption{(Color online) (a) $\alpha^{\mathrm{int}}$, (b) $\beta^{\mathrm{int}}$, and (c) $\gamma^{\mathrm{int}}$ as functions of $E$ and $X$.}
\label{fig:abg}
\end{figure}

It is convenient to define the intrinsic polarizability as
\begin{equation}
\alpha^{\mathrm{int}}\left(X,E\right) = \frac{\alpha\left(X,E\right)}{\alpha^{\mathrm{max}}} = \left[X^2+E^2 \left(1-X^2\right)\right]  ,
\label{eq:IntrinsicPol}
\end{equation}
which is bound by $0 \leq \alpha^{\mathrm{int}}\left(X,E\right) \leq 1$.  Thus, $\alpha^{\mathrm{int}}\left(X,E\right)$ is a measure of the polarization as a fraction of the fundamental limit.  More importantly, one can show that the intrinsic polarizability is a scale invariant quantity that allows one to compare molecules of different sizes.\cite{kuzyk10.01}  As such, the results of our calculations are applicable to molecules of all sizes.

The top graph in Figure \ref{fig:abg} shows the intrinsic polarizability, $\alpha^{\mathrm{int}}$, as a function of $X$ and $E$.  The intrinsic polarizability is a maximum when $X = 1$.

To calculate the higher-order susceptibilities, the sum rules are used to determine the other moments in terms of $E$ and $X$:
\begin{eqnarray}
|x_{12}| &=& |x_{10}^{\mathrm{max}}|\sqrt{\frac{E}{1-E}}\sqrt{1+X^2} ,\label{eq:x12}  \\
|x_{10}|\Delta x_{10} &=& |x_{10}^{\mathrm{max}}|^2 \frac{E-2}{\sqrt{1-E}}\sqrt{1-X^4} ,\label{eq:x10Dx10} \\
|x_{20}|\Delta x_{20} &=& |x_{10}^{\mathrm{max}}|^2\left(1-2E\right)\sqrt{\frac{E}{1-E}}X\sqrt{1+X^2} , \nonumber \\
\label{eq:x20Dx20}
\end{eqnarray}
where $\Delta x_{10}=x_{11}-x_{00}$ and $\Delta x_{20}=x_{22}-x_{00}$.  The hyperpolarizability can be expressed in terms of $E$ and $X$ by substituting Equations \ref{eq:Edefine}-\ref{eq:finalx10x20} and \ref{eq:x12}-\ref{eq:x20Dx20} into Equation \ref{eq:betasum},\cite{kuzyk09.01} thereby giving
\begin{eqnarray}
\beta_{3L} &=& 6 e^3 \frac{|x_{10}^{\mathrm{max}}|^3}{E_{10}^2}
\left(1-E\right)^{3/2} \nonumber \\
&\times& \left(E^2+\frac{3}{2}E+1\right)X\sqrt{1-X^4} .
\label{eq:finalbeta}
\end{eqnarray}
Equation \ref{eq:finalbeta} can be rewritten using Equation \ref{eq:x10max} to give the intrinsic hyperpolarizability
\begin{eqnarray}
\beta^{\mathrm{int}} \left(X,E\right) &=& \frac{3^{3/4}}{\sqrt{2}} \left(1-E\right)^{3/2} \nonumber \\ &\times& \left(1+\frac{3}{2}E+E^2\right)X\sqrt{1-X^4} .
\label{eq:betaN}
\end{eqnarray}
Figure \ref{fig:abg}b shows the intrinsic hyperpolarizability, $\beta^{\mathrm{int}}$, as a function of $X$ and $E$.  When $E=0$ and $X = \sqrt[-4]{3}$, $\beta^{\mathrm{int}}=1$.

The intrinsic second hyperpolarizability for a three level model can be written by substituting Equations \ref{eq:x10max}-\ref{eq:finalx10x20} and Equations \ref{eq:x12}-\ref{eq:x20Dx20} into Equation \ref{eq:gammasum} and summing over the first two excited states. This leads to an intrinsic second hyperpolarizability,\cite{kuzyk00.02,perez01.08}
\begin{eqnarray}
\gamma^\mathrm{int} \left(X,E\right) &=& \frac{1}{4} \left[4 - 2(E^2-1)E^3 X^2 \right. \nonumber \\
&-& 5 \left(E-1\right)^2\left(E+1\right)\left(E^2+E+1\right)X^4 \nonumber \\
&-& \left. \left(E^3+E+3\right)E^2 \right].
\label{eq:gammaN}
\end{eqnarray}
Figure \ref{fig:abg}c shows the intrinsic second hyperpolarizability, $\gamma^\mathrm{int}$, as a function of $X$ and $E$. Note that $\gamma^\mathrm{int} = 1$ when $X=0$ and $E=0$ and $\gamma^\mathrm{int} = - 1/4$ when $E=1$.

\section{Results and Discussion}

\subsection{Effective Susceptibilities in the weak interaction limit}
\label{sec:cascade3L}

Equations \ref{eq:IntrinsicPol}, \ref{eq:betaN}, and \ref{eq:gammaN} relate the intrinsic molecular susceptibilities to $X$ and $E$. Using these results, we can determine the first three effective molecular susceptibilities, which are of the form $k^{\left(n\right)}_\mathrm{eff}\left(r,X,E\right)$. With $\alpha\left(X,E\right) = \alpha^{\mathrm{max}} \alpha^{\mathrm{int}} \left(X,E\right) $, $\beta\left(X,E\right) = \beta^{\mathrm{max}} \beta^{\mathrm{int}} \left(X,E\right) $, and $\gamma\left(X,E\right) = \gamma^{\mathrm{max}} \gamma^{\mathrm{int}} \left(X,E\right) $ substituted into Equations \ref{eq:aeff} - \ref{eq:geff}, we obtain an equation for the first three effective molecular susceptibilities as a function of $r$, $X$, and $E$, where
\begin{eqnarray}
\alpha_{\mathrm{eff}}\left(X,E,r\right) &=& \frac{2r^3 \alpha\left(X,E\right)}{r^3 + \alpha\left(X,E\right)} , \label{eq:beginalpha} \\
\beta_{\mathrm{eff}}\left(X,E,r\right) &=& \frac{2r^9 \beta\left(X,E\right)}{\left(r^3 + \alpha\left(X,E\right)\right)^3} , \label{eq:beginbeta} \\
\gamma_{\mathrm{eff}}\left(X,E,r\right) &=& \frac{2r^{12}\gamma\left(X,E\right)}{\left(r^3 + \alpha\left(X,E\right)\right)^4} \nonumber \\  \label{eq:begingamma}
&-& \frac{4 r^{12} \left[\beta\left(X,E\right)\right]^2}{\left(r^3 + \alpha\left(X,E\right)\right)^5} .
\end{eqnarray}

It is important to point out that Equations \ref{eq:beginalpha}-\ref{eq:begingamma} show that the polarizability and hyperpolarizabilities can not be independently varied, but are parameterized by $E$ and $X$.  As we show below, it is not permissible to simultaneously make the direct second hyperpolarizability large when the cascading contributions are also large.

We first treat the case where the molecules do not interact so that $N$ and $E_{10}$ do not change as a function of $r$.  In this case, the intermediate cascading photon is the only link between the two molecules.  We call this the non-interacting case because the molecular states remain unperturbed. Then, the fundamental limit of the (hyper)polarizabilities is simply the sum of the individual values of each molecule.  As we describe later, this approximation overestimates the cascading contribution because it underestimates the fundamental limits.

In the non-interacting limit, the intrinsic effective molecular susceptibilities as a function of $X$ and $E$ are
\begin{eqnarray}
\alpha_\mathrm{eff}^\mathrm{int}\left(X,E,r\right) = \frac{\alpha_\mathrm{eff}\left(X,E,r\right)}{2\alpha^\mathrm{max}} , \label{eq:ainteff} \\
\beta_\mathrm{eff}^\mathrm{int}\left(X,E,r\right) = \frac{\beta_\mathrm{eff}\left(X,E,r\right)}{2\beta^\mathrm{max}} , \label{eq:binteff} \\
\gamma_\mathrm{eff}^\mathrm{int}\left(X,E,r\right) = \frac{\gamma_\mathrm{eff}\left(X,E,r\right)}{2\gamma^\mathrm{max}} , \label{eq:ginteff}
\end{eqnarray}
where the factor of two in the divisor results from the additivity of susceptibilities.

In gaussian units, the polarizability has units of volume.  In rough terms, the fundamental limit of the polarizability $\alpha_{\mathrm{max}}$ defines the largest possible polarization volume, which represents a length scale $\sqrt[3]{\alpha_{\mathrm{max}}}$.  Using Equation \ref{eq:alphaMAX}, we can express the separation of the molecules in terms of this natural length scale,
\begin{equation}
r^\prime = r/\sqrt[3]{\alpha_{\mathrm{max}}} =  \left(\frac{m E_{10}^2}{e^2 \hbar^2 N}\right)^{\frac{1}{3}} r . \label{eq:rprime}
\end{equation}

\subsection{Effective second hyperpolarizability and cascading}

\begin{figure}[t!]
\centering\includegraphics{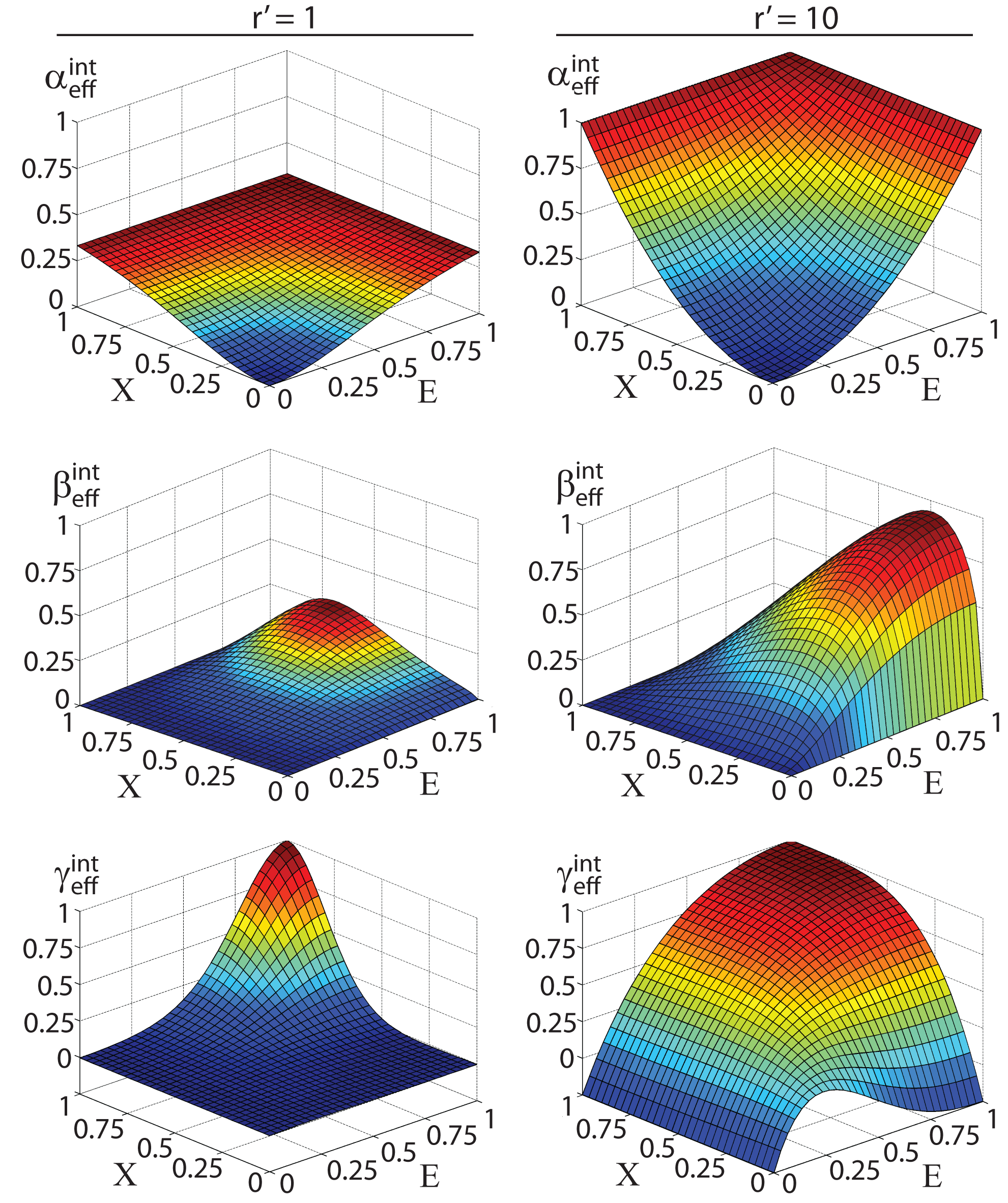}
\caption{(Color online) The effective molecular susceptibilities $\alpha_{\mathrm{eff}}\left(X,E,r\right)$, $\beta_{\mathrm{eff}}\left(X,E,r\right)$, and $\gamma_{\mathrm{eff}}\left(X,E,r\right)$, when the distance of separation is $r^\prime=1$ and $r^\prime=10$.}
\label{fig:gint0}
\end{figure}

The intrinsic effective molecular susceptibilities from cascading -- $\alpha_{\mathrm{eff}}^{\mathrm{int}}$, $\beta_{\mathrm{eff}}^{\mathrm{int}}$, and $\gamma_{\mathrm{eff}}^{\mathrm{int}}$ -- as a function of $X$ and $E$ at $r^\prime = 1$ and $r^\prime = 10$ are shown in Figure \ref{fig:gint0}.  When $r^\prime =1$, the molecules are close together and cascading is large while when $r'=10$, the molecules are much further apart than their sizes, so cascading should be negligible.  A comparison between Figure \ref{fig:abg} and the righthand portion of Figure \ref{fig:gint0}, for which $r^\prime = 10$, shows that for $r^\prime \rightarrow \infty$, the cascading calculation reduces to the same result as one gets for two independent molecules.

A comparison of the left and righthand side of Figure \ref{fig:gint0}, which represents large and small contributions to cascading, leads to several conclusions.  First,  the effective second-order hyperpolarizability, $\gamma_{\mathrm{eff}}$, never exceeds the fundamental limit.  Thus, cascading does not provide a loophole that allows the largest possible value of $\gamma_{\mathrm{eff}}$ to be larger than the fundamental limit of two individual molecules.

The difference $\gamma_{\mathrm{eff}}\left(E,X\right) - \gamma_{\mathrm{direct}} \left(E,X\right) \approx \gamma_{\mathrm{eff}} \left(X,E,r^\prime = 1 \right) - \gamma_{\mathrm{eff}}\left(X,E,r^\prime =10\right)$ is an estimate of the cascading contribution.  When the direct second hyperpolarizability peaks at the fundamental limit (at $E=0$ and $X=0$), the cascading contribution vanishes, as we would expect if the fundamental limit is to be obeyed.  However, cascading in the current geometric configuration interferes with the direct contribution near the peak, leading to a smaller second hyperpolarizability in the vicinity of the peak.

The direct second hyperpolarizability's negative fundamental limit is given by $\gamma^{\mathrm{int}} = -0.25$.  In the regions where $\gamma^{\mathrm{int}} \approx -0.25$, the cascading contribution cancels the direct term, leading to a vanishing second hyperpolarizability.  Thus, side-by-side cascading appears to be a nuiscence rather than a design strategy for making large-$\gamma$ systems.

Note that the true separation distance, $r$, is related to the dimensionless separation $r^\prime (E_{10},N)$, which depends on the number of electrons and transition energy to the first excited state.  Molecules with a greater number of electrons have a larger effective size, and therefore give a larger cascading contribution for a fixed separation than molecules with fewer electrons.

\subsection{Molecular design using cascading}

The fact that there exist examples where the cascading contribution is larger that the direct contribution is not at odds with our results.  The salient point is that molecules that have small second hyperpolarizabilities may have a relatively large cascading contribution.  However, given a fixed pallet of building blocks, i.e. electrons and nuclei, an optimized arrangement of nuclei\cite{kuzyk06.02} will always yield a larger second hyperpolarizability than what one can attain with cascading between two separate molecules with the same total number of electrons.  Thus in principle, it is always better to design a large molecule than splitting it up into two smaller ones with the same number of electrons and using cascading.

The crucial issue, however, is that nature may not allow nuclei to be placed into the ideal geometry. Indeed, that may be one of the reasons for the large gap between the best molecules and the fundamental limit.\cite{Tripa04.01,zhou08.01,kuzyk09.01,kuzyk10.01} If that is the case, cascading may offer an additional degree of design flexibility as described by Baev and coworkers.\cite{baev10.01}  To investigate the promise of such an approach, we consider $\gamma_{\mathrm{eff}}^{\mathrm{int}} \left( X , E \right)$ when cascading dominates, which is the case when the direct second hyperpolarizability vanishes, or  $\gamma^{\mathrm{int}}\left(X,E\right) = 0$.

\begin{figure}
\centering\includegraphics{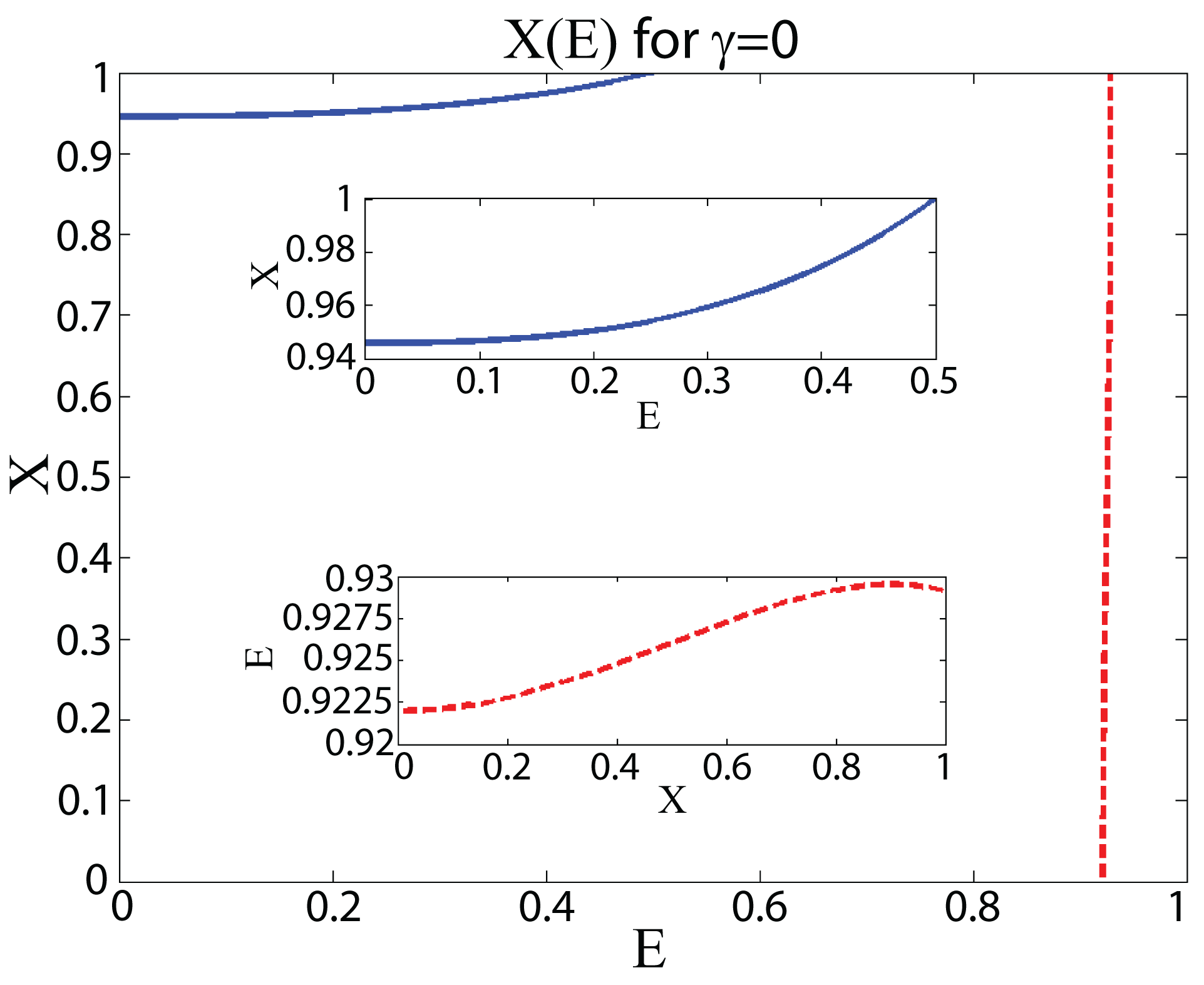}
\caption{(Color online) The function $X(E)$ when $\gamma = 0$ for $0\leq X \leq 1$ and $0\leq E \leq 1$. The insets show an expanded view.}
\label{fig:XofE}
\end{figure}

Figure \ref{fig:XofE} shows a plot of the domain $(E,X)$ where $\gamma^{\mathrm{int}} \left(X , E \right) = 0$. The blue solid curve is represented by the function
\begin{equation}
X\left(E\right) = \frac{1}{\sqrt{5}}\left(\frac{E^3 - E^5 + \sqrt{f\left(E\right)}}{1 - E^2 - E^3 + E^5}\right)^{\frac{1}{2}} ,
\label{eq:xfunctione}
\end{equation}
where
\begin{eqnarray}
f\left(E\right) &=& 20 - 35 E^2 - 25 E^3 + 15 E^4 + 35 E^5\nonumber \\
&+& 6 E^6 - 10 E^7 - 2 E^8 - 4 E^{10}
\label{eq:fofe}
\end{eqnarray}
for $0\leq X \leq 1$. The red dotted curve was found numerically because $E\left(X\right)$ is a high degree polynomial, and the root could not solved analytically.

\begin{figure}
\centering\includegraphics{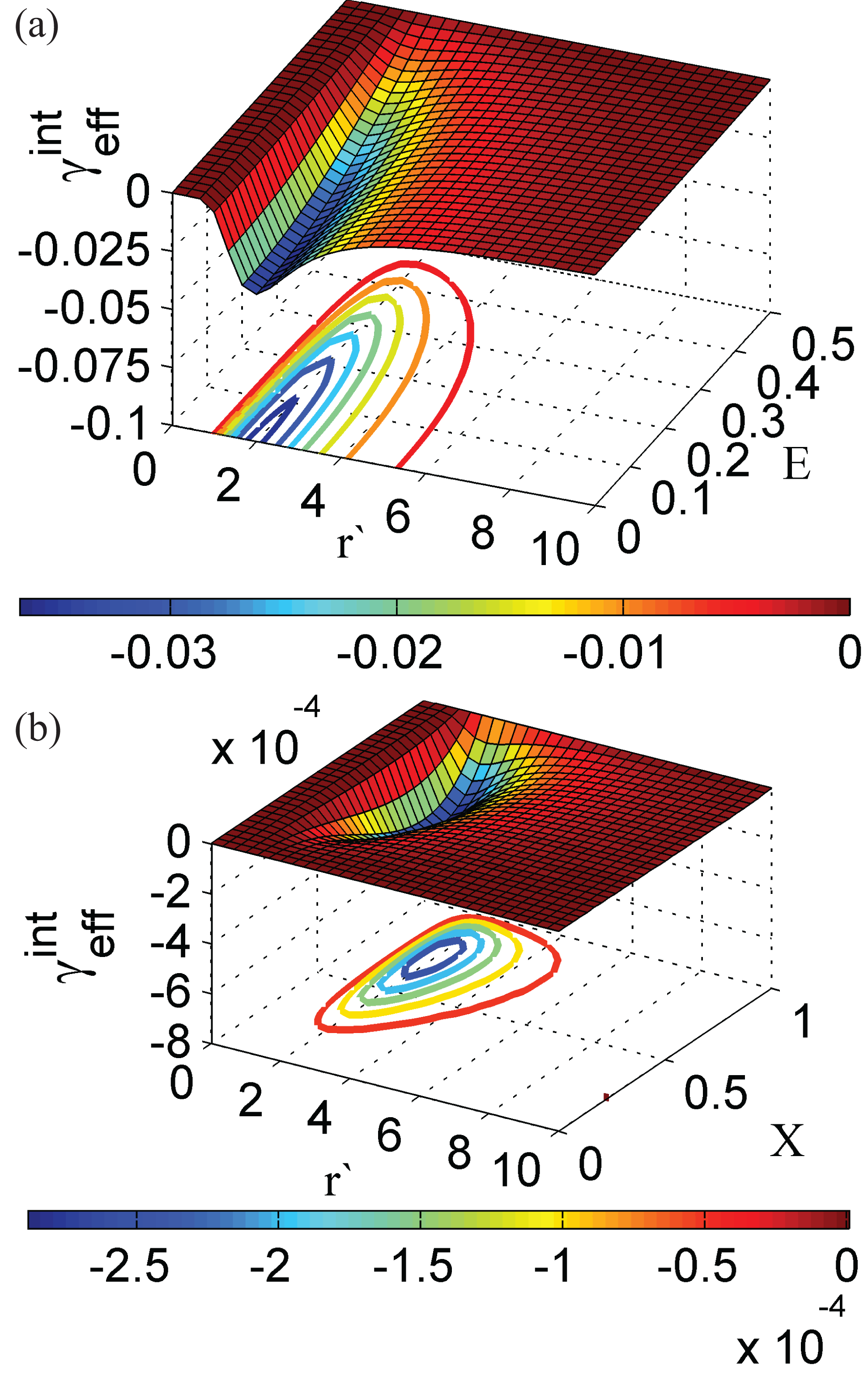}
\caption{(Color online) (a) A surface plot of $\gamma_{\mathrm{eff}}^{\mathrm{int}}\left(r^\prime,E\right)$ along the branch corresponding to the solid blue curve in Figure \ref{fig:XofE}; and, (b) $\gamma_{\mathrm{eff}}^{\mathrm{int}}\left(r^\prime,X\right)$, along the dashed red curve.}
\label{fig:g0XE}
\end{figure}

Along the curve $X(E)$ where the direct second hyperpolarizability vanishes,  $\gamma_{\mathrm{eff}}^{\mathrm{int}}$ can be expressed as a function of either $r^\prime$ and $X$, or $r^\prime$ and $E$. Figure \ref{fig:g0XE}a shows $\gamma_{\mathrm{eff}}^{\mathrm{int}}$ along the blue solid curve $X\left(E\right)$ in Figure \ref{fig:XofE} as a function of $r^\prime$. Figure \ref{fig:g0XE}b shows the same plot but along the red dashed curve $E\left(X\right)$ in Figure \ref{fig:XofE}

The cascading contribution shown in Figure \ref{fig:g0XE}a due to the blue solid branch peaks at energy ratio $E=0$ and normalized separation $r^\prime \approx 2$, and contributes just over -0.035 to the intrinsic second hyperpolarizability - a factor of almost 30 below the fundamental limit for a direct third-order susceptibility. The red branch, shown in Figure \ref{fig:g0XE}b, contributes less than $3 \times 10^{-4}$ to the intrinsic second hyperpolarizability.  There are other regions in the $(E,X,r^\prime)$ domain where the cascading contribution is larger; but, it is of opposite sign to the direct contribution, and makes the effective third-order susceptibility smaller than if cascading were absent.

To summarize, we find three distinct regions:

\begin{enumerate}

\item{The direct contribution is large and the cascading contribution is negligible.  This leads to the largest second hyperpolarizability.}

\item{The direct contribution vanishes and only cascading contributes.  This leads to intrinsic second hyperpolarizabilities that are more than a factor of 30 smaller than in case 1.}

\item{The cascading contribution is large.  In this regime, cascading cancels the direct term, leading to small nonlinear susceptibilities.}

\end{enumerate}

Thus, the {\em largest possible nonlinear response} is in the limit when there is no cascading contribution for the side-by-side configuration.  We emphasize that this does not imply that the cascading contribution can not be larger than the direct contribution.  However, in these cases the second hyperpolarizability will fall far short of the fundamental limit.

\subsection{Strongly interacting molecules}

In the previous section, we considered only weak interactions, where the wave functions of each molecule were not affected by their proximity to the other molecule.  In the limit of weak interaction, the fundamental limit of the susceptibilities of a two-molecule system is simply the sum over the fundamental limits of each molecule.  However, in the case of strong interactions, the two molecules can be viewed as one, where cascading is one of many electromagnetic interactions between what we would consider the two separate molecules.  Clearly, if the cascading interactions are as strong as the interaction between the electrons in each molecule, then each molecule loses its separate identity.

The fundamental limits of the off-resonant susceptibilities can be written as,
\begin{equation}\label{alphamax}
\alpha_0^{\mathrm{max}} = \left( \frac {e \hbar} {\sqrt{m}} \right)^2 \frac {N} {E_{10}^2},
\end{equation}
\begin{equation}\label{betamax}
\beta_0^{\mathrm{max}} = \sqrt[4]{3} \left( \frac {e \hbar} {\sqrt{m}} \right)^3 \left[ \frac {N^{3/2}} {E_{10}^{7/2}} \right] ,
\end{equation}
and
\begin{equation}\label{gammamax}
-\left( \frac {e \hbar} {\sqrt{m}} \right)^4 \frac {N^2} {E_{10}^5} \leq \gamma_0 \leq 4 \left( \frac {e \hbar} {\sqrt{m}} \right)^4  \frac {N^2} {E_{10}^5} \equiv \gamma_0^{\mathrm{max}}.
\end{equation}

Equations \ref{alphamax} through \ref{gammamax} can be used to determine the scaling properties of various quantities.  For example, the fundamental limit of the polarizability is equivalent to the upper limit of the cube of the the radius, or $\left(r_0/2\right)^3 \sim \alpha_0^{\mathrm{max}}$. This defines a fundamental limit of the polarization length of $x_0 \sim \sqrt[3]{\alpha_0^{\mathrm{max}}}$.  For a hydrogen atom, $E_{10} = 10.2\,eV$ and N=1 yielding $x_0 = 1 \AA$, the Bohr diameter.

The concept of the polarization breaks down when $\gamma E^2 \sim \alpha$.  Since $\gamma$ and $\alpha$ are maximal under the same conditions in the three-level ansatz, the expression for the polarization becomes meaningless when,
\begin{equation}\label{MaxField}
E_0 \sim \frac {1} {2} \left( \frac  {\sqrt{m}} {e \hbar} \right) \frac {E_{10}^{3/2}} {N^{1/2}} .
\end{equation}
Using the hydrogen atom for illustration, Equation \ref{MaxField} gives a characteristic field limit that is smaller (but within an order of magnitude) than the electric field experienced by the electron.\footnote{Note that if we had used $\beta E \sim \alpha$ to calculate the limit of the characteristic field strength, the result would have been 50\% greater than what is given by Equation \ref{MaxField} -- a good ballpark estimate.  However, since $\beta$ vanishes in a centrosymmetric systems, using $\alpha$ and $\gamma$ for determining scaling rules is safer.}

First, we consider the fundamental limit of the direct second hyperpolarizability for two molecules.  If each atom has $N$ electrons, and the fundamental limit for one molecule is $\gamma_0^{\mathrm{max}}$, then the fundamental limit for two noninteracting electrons is $2\gamma_0^{\mathrm{max}}$.  If the two molecules are combined so that the electrons fully mix, the fundamental limit is $4\gamma_0^{\mathrm{max}}$ provided that the transition energy $E_{10}$ remains the same.  When $N$ molecules are combined, the direct second hyperpolarizability scales as $N$ while combining the molecules into one super molecule leads to $N^2$ scaling.  Thus, it is best to make a large molecule rather than many small ones.  Similarly, it may be better to allow these molecules to interact with each other.

As an illustration, we calculate the scaling behavior of the fundamental limits of a system with energy-level spacing like a polyene molecule and compare it with the model of a polyene of Rustagi and Ducuing.\cite{rusta74.01}  Hans Kuhn showed that light absorption in polyenes could be modeled by an infinite box filled with $N$ electrons that obey the pauli exclusion principle, where $N/2$ is the number of double bonds in the conjugated chain.\cite{kuhn48.01,kuhn49.01}

We consider two cases. If $n$ is an even state, these molecules are called polyenes. The ground state consists of two electrons in each energy level up to and including the state $n=N/2$.  The first excited state will correspond to promoting an electron from state $n=N/2$, the highest occupied energy level, to the state $n=N/2 +1$, the lowest unoccupied level.  The transition energy is then given by,
\begin{equation}\label{EvenElectronExcitation}
E_{10}=\frac{\hbar^2\pi^2}{2mL^2}\left(\left[ \frac {N} {2} + 1 \right]^2 - \left[ \frac {N} {2} \right]^2 \right) = \frac{\hbar^2\pi^2}{2mL^2}\left(N+1 \right),
\end{equation}
where $L$ is the length of the molecule.  The second excited state corresponds to promoting the remaining electron from the first excited state in level $n=N/2$ to level $n=N/2 +1$, and yields $E_{20} = 2 E_{10}$

For an odd number of electrons, such molecules are called cyanines.  All states through $n=(N-1)/2$ are each occupied with two electrons and the state with $n=(N-1)/2 + 1$ is singly occupied.  Thus, the lowest excited state energy corresponds to promoting an electron from the state with $n=(N-1)/2$ to the state with $n=(N-1)/2 + 1$.  This leads to
\begin{equation}\label{OddElectronExcitation}
E_{10} = \frac{\hbar^2\pi^2}{2mL^2}N.
\end{equation}
The second excited state is formed by promoting an electron from the system's ground state from state $n=(N-1)/2$ to the state with $n=(N-1)/2 + 2$.  In this case, $E_{20} = \frac {2(N+1)} {N} E_{10}$.  Thus for small $N$, $E_{20}/E_{10}$ is large.  Note that in the limit of large $N$, Equations \ref{EvenElectronExcitation} and \ref{OddElectronExcitation} converge, as do the values of the ratio $E_{20}/E_{10}$.  Thus, we will us Equation \ref{OddElectronExcitation} in the rest of the analysis below.

Defining the average linear electron density $\lambda = N/L$ and the Bohr radius $a_0 = \hbar^2/me^2$, the energy of Equation \ref{OddElectronExcitation} can be expressed as,
\begin{equation}\label{OddElectronExcitationPolyene}
E_{10} = \frac{\pi^2 e^2 a_0 \lambda}{2L}  ;
\end{equation}
and, the susceptibilities in Equations \ref{alphamax} through \ref{gammamax} in the large $N$ limit for a polyene become,
\begin{equation}\label{alphamaxPolyene}
\alpha_{\mathrm{poly}}^{\mathrm{max}} = \frac {4 } {\pi^2 a_0 \lambda} L^3,
\end{equation}
\begin{equation}\label{betamaxPolyene}
\beta_{\mathrm{poly}}^{\mathrm{max}} =   \frac { 2^{7/2} \sqrt[4]{3} } {\pi^7 \lambda^2 e a_0^2}  L^5 ,
\end{equation}
and
\begin{equation}\label{gammamaxPolyene}
\gamma_{\mathrm{poly}}^{\mathrm{max}} = \frac {128} {\pi^{10} e^2 a_0^3 \lambda^3} L^7 .
\end{equation}

The model of Rustagi and Ducuing gives,\cite{rusta74.01}

\begin{equation}\label{alphamaxPolyeneRD}
\alpha_{\mathrm{poly}}^{\mathrm{RD}} = \frac {8 } {3 \pi^2 a_0 \lambda} L^3,
\end{equation}

\begin{equation}\label{gammamaxPolyeneRD}
\gamma_{\mathrm{poly}}^{\mathrm{RD}} = \frac {256 } {45 \pi^6 e^2 a_0^3 \lambda^5} L^5 .
\end{equation}

One may argue that the cascaded second hyperpolarizability might scale more favorably than the direct one. It is simple to show that both scale in the same way.  Cascading in the side-by-side configuration results from the second term in Equation \ref{eq:geff}.  Since the molecules cannot be closer together than approximately the characteristic size, which at the fundamental limit is given by $\sqrt[3]{\alpha_0^{\mathrm{max}}}$, the cascading term scales as
\begin{equation}\label{MaxField}
\gamma_{\mathrm{cascading}}^{\mathrm{max}} = \frac {\left( \beta_0^{\mathrm{max}} \right)^2} {\alpha} = \sqrt{3} \left( \frac {e \hbar} {\sqrt{m}}  \right)^4 \frac {N^2} {E_{10}^{5}} \sim \gamma_{\mathrm{direct}}^{\mathrm{max}} ,
\end{equation}
where we have used the fundamental limits in Equations \ref{alphamax} through \ref{gammamax} to determine how the susceptibilities scale when the hyperpolarizability is optimized, a criteria for optimizing $\beta^2$.  Hence, the cascading term scales in the same way as the direct second hyperpolarizability.

\subsection{Linear local electric fields}

The self-consistent field calculation used to derive Equations \ref{eq:aeff}-\ref{eq:geff} are, in effect, calculating the local electric field, $f^0$.  Cascading is the lowest-order local field correction to the hyperpolarizability.  It is important to stress that this calculation also treats the effective linear local electric field (i.e. originating in the polarizability) due to the neighboring molecule.

From Equations \ref{eq:aeff} through \ref{eq:geff}, it is clear that the effective linear local electric field, $f^0 = E_a + E_{\mathrm{loc}} = E_a/\left(1+\alpha/r^3\right)$ where $E_a$ is the applied field, scales as
\begin{equation}\label{LocalLinearField}
\frac{f^0}{E_a} = \frac {1} {1+\alpha/r^3} ,
\end{equation}
and the net polarization field scales as
\begin{equation}
\frac{E_{\mathrm{loc}}}{E_a} = \frac {-1} {r^3/\alpha+1} .
\label{eq:localpolfield}
\end{equation}

\subsection{Analysis of Approximations}

It is worthwhile to summarize the various approximations used and whether or not they restrict the applicability of the results to real molecules.

\subsubsection{The dipole approximation and the point dipole}

Most of nonlinear optics is based on the dipole approximation, and is embodied in the expansion of the polarization in a power series of the electric field.  This approximation holds when the dipole coupling energy is larger than the energy of higher-order moments.  The quadrupole moment couples to the electric field gradient, the octupole moment to the a dyad of field gradients, etc.  As long as the electric field does not vary much on the scale of the size of a molecule, the dipole approximation is appropriate - a condition that holds well when the wavelength of the light is long compared with the size of the molecule.

Near interfaces, the electric field can change abruptly, rendering the dipole approximation invalid.  An important question is whether or not the variation of the electric field is large enough in the vicinity of a molecule to require higher-order moments to be taken into account.

For two isolated and stationary molecules, the field gradients near each one can be large.  However, in real systems with many molecules, large local field fluctuations cancel out and the local fields due to the surrounding material can be approximated as smooth.  If this were not the case, the Lorentz Lorenz local field models and the Onsager models would not provide reasonable descriptions of a molecule in a medium.  Thus, using the dipole approximation to describe pair interactions, then summing over all possible pairs, removes, on average, the higher-order contributions.  Indeed, this process of averaging can be shown to be equivalent to the physics underlying cavity field models.  Thus, the pair interaction of two dipoles is the appropriate fundamental quantity for any cascading calculation.

One can rightly argue that higher-order terms, as are required for cascading, do not lead to a linear smoothing effect because the average over an electric field to a power $n$ is not equal to the average over the $n$th power of the electric field.  As such, it is important that fluctuations in the electric field not be too large.  This will be the case if the separation between molecules is larger than the size of charge density fluctuations within a molecule.  As we show below, this approximation may hold even when the molecules touch.

The point dipole approximation for an ideal dipole holds when the separation between dipoles is much larger than separation between the two point charges in each dipole.  In a molecule, this roughly translates into the separation between molecules exceeding their size.  The dipole moment of a molecule is usually much smaller than the product of its size and the number of electrons.  Thus, one can picture a molecule as being composed of a large number of small dipoles whose moments are determined from fluctuations of the wave functions over short distances.  Thus, the calculations may hold even when the molecules are close to touching.

When the molecules get too close together, the $1/r^3$ dipole form of the field has higher-order  correction terms in powers of $a/r$. where $a$ is the size of the dipole moment.  Interestingly, the next higher-order correction factor is negative, and thus decreases the strength of interactions, and therefore decreases the strength of cascading.  Since we are interested in the upper limits enhancement of the second hyperpolarizability upon using cascading, neglect of the lowest-order corrections overestimates the cascading contribution.

Once the separation between centers of the molecules is less than the electromagnetic size of the molecule, it no longer makes sense to consider the system as two distinct molecules.  In this limit, the intermolecular coulomb forces become comparable to the intramolecular coulomb forces, so the individual character of each molecule is lost, and the two together act as a single quantum system; that is, a single molecule.  For these separations, the cascading calculations break down because the meaning of cascading breaks down.

\subsubsection{One-dimensional approximation}

For simplicity, all of our models assume that the largest tensor component dominates all others; or, that only the largest tensor component is used in the process.  The 1D approximation is extensively used, especially in the study of donor-acceptor molecules, donor-donor molecules and acceptor-acceptor molecules.\cite{kuzyk98.01} While this approximation has been highly successful in the study of a large number of molecules, it would be useful to check how this may impact the cascading calculations.

Table \ref{table:geometry} illustrates the cascading third-order polarization for different molecular arrangements. Clearly, if the applied electric field at molecule $a$ and the field from $p^{(2)}$ at molecule $a$ due to molecule $b$ are aligned along the largest tensor element of $\beta$ of molecule $a$; and if the same holds for molecule $b$ - i.e. the arrangements of the molecules are reciprocal, than the cascading effect will be optimized.  The simplest case is shown in the first column of Table \ref{table:geometry}, where the molecules are in the end-to-end arrangement with the applied field and molecular fields in the same direction.  Thus, the cascading term will be proportional to $\beta_{xxx}^2$.  This case will be described in detail in the companion article.\cite{dawson11.02a}

\begin{table}[b!]
\caption{The cascading third-order polarization, $ p_{x}^{(3)}$, for the arrangement shown in each diagrams.}
\centering
\begin{tabular}{c c c}
\hline\hline
 End-to-End & Side-by-Side & General \\
 $\beta_{xxx}^2 E^2 E_a$ & $ \beta_{yxx}^2 E^2 E_a$ & $ \sum_{i} \beta_{ixx} \beta_{ixx} E^2 E_i$ \\[2pt]
\hline \\[-1.1em]
\includegraphics{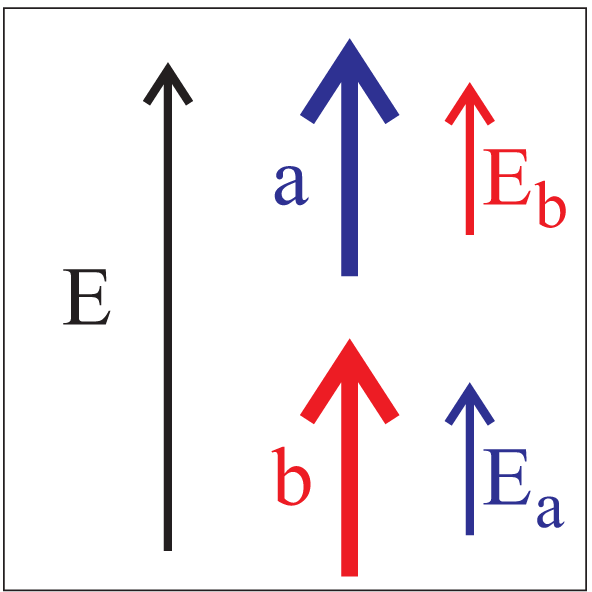} & \includegraphics{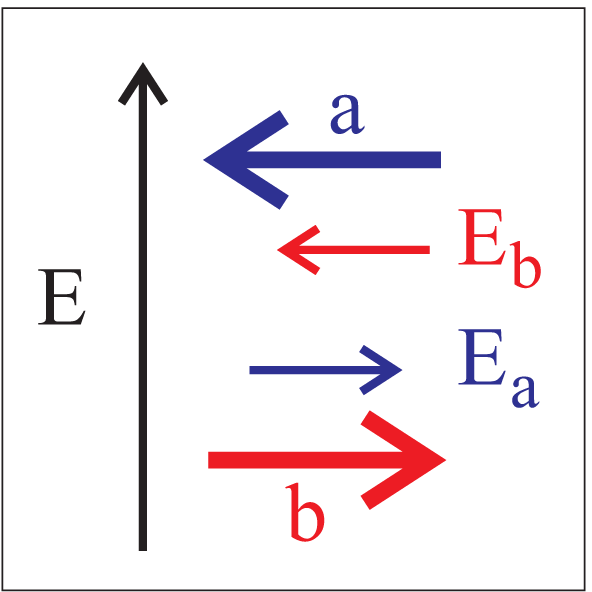} & \includegraphics{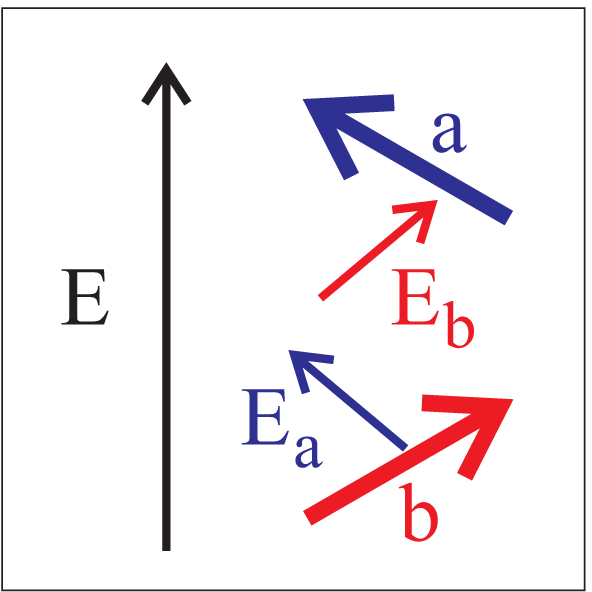} \\
\hline
\end{tabular}
\label{table:geometry}
\end{table}

As we have shown in this paper, the side-by-side case with one-dimensional molecules yields a vanishing cascading term when they are counteraligned. However, when the molecules are two-dimensional, so that they have a nonvanishing component of $\beta_{yxx}$, then two applied fields perpendicular to the long axis of each molecule will yield a second-order dipole moment along the long axes that leads to a dipole field that acts along the dipole axis of each of the other molecules, as shown in Table \ref{table:geometry}.  Thus, the cascading term will be proportional to $\beta_{yxx}^2$.  Inasmuch as this tensor component is comparable to the diagonal one, the cascading contribution will be half as large due to the geometric factor in the dipole field.

The third column in Table \ref{table:geometry} shows the general case where the applied fields and induced dipole fields are in arbitrary directions.  This configuration will yield the largest cascading term if the applied electric field yields a second-order induced dipole moment whose electric field at the second dipole moment, when properly oriented, reinforces the nonlinear interaction between the applied electric field and the dipole field.  For example, when only the three tensor components $\beta_{ixx}$ for $i = 1,2,3$ are nonzero, this cascading term will be of the form $ \sum_{i} \beta_{ixx} \beta_{ixx} E^2 E_i$, where $E_i$ is the dipole field along direction $i$.  Since this sum can be no larger than the dipole field times the largest tensor component of $\beta$, this case cannot yield a value that is larger than the case for the one-dimensional molecule.

In using the one-dimensional approximation and aligning the molecules in a way that optimizes the cascading term, one would expect to get the largest possible cascading contribution.  Since our interest is in understanding the magnitude of the largest attainable response, the one-dimensional approximation does not limit the generality of the result.  However, taking into account the relevant tensor components is clearly critical in modeling a particular system.

\subsubsection{Three-level Ansatz}

The three-level ansatz (TLA) states that when a quantum system has a nonlinear response that is at the fundamental limit, it is described by a three-level model.\cite{zhou07.02,zhou08.01,kuzyk09.01}  While there is no analytical proof, the TLA has been shown to hold in all cases ever studied.\cite{kuzyk09.01,shafei10.01,shafe11.01} However, for systems with a nonlinear response substantially smaller than the limit, more states can contribute.  Thus, our results can be highly uncertain in these cases.  However, when the nonlinear response is large, the regime of interest, the calculations should be more accurate.

\subsubsection{Summary of the Approximations}

The dipole, point particle, and 1D approximations serve to understand the largest possible response that is attainable with cascading.  While an equivalently large response is possible in real systems where the 1D approximation is relaxed, the point particle and dipole approximations should approximately hold for most real systems.  However, if the dipole approximation were strongly violated, then the theory presented here would fail, but so would a description of the nonlinear optical properties of a material in terms of linear and higher order hyperpolarizabilities.

Though we treat only special arrangements of the dipoles, these are the ones that are expected to give the largest results.  The companion article discusses more realistic systems in which the orientations are averaged as they would be in a liquid or dye-doped polymer.

\section{Summary}
\label{Summary}

We have applied a self-consistent field calculation to determine the induced dipole moments of two molecules in close proximity in the presence of an applied electric field.  From these, we use a semiclassical approach to determine the nonlinear response of the dipole pair using the quantum-derived sum-over-states expressions for the direct polarizability and hyperpolarizability to determine the cascading contribution to the second hyperpolarizability.  The semi-classical approximation assumes that the energies of the system are not affected by the interaction, leading to an overestimate of the cascading contribution.

The sum rules and the three-level ansatz are applied to each molecule, thus decreasing the number of parameters that characterize the system to two parameters: $E$, the energy ratio; and, $X$, the normalized transition moment.  In this way, the cascading and direct terms can be compared over the full domain spanned by $E$ and $X$.

In order to remove scaling effects, the intrinsic nonlinear-optical response is used for comparison.  The intrinsic values are calculated by normalizing the results to the fundamental limit.  Then, the intrinsic second hyperpolarizability of the molecule pair is calculated as a function of the normalized transition moment, $X$, the energy ratio, $E$, and the distance between molecules, $r$, for the aligned and counter-aligned cases.  Counter-aligned molecules lead to a null result.

The largest effective intrinsic second hyperpolarizability - the sum of the direct second hyperpolarizability and cascading contribution - are independent of $r$ with $\gamma_{\mathrm{eff}}^{\mathrm{int}} =1$.  However, for each $r$, this peak is found at a unique point in the domain spanned by $(X,E)$.  Thus, when the direct value of $\gamma$ is optimized the cascading contribution vanishes, while a large cascading contribution demands that the direct contribution be sub-optimal.  Thus, the cascading process does not provide a back door for breaking the fundamental limits.  Interestingly, when $\gamma$ of the direct contribution vanishes, the cascading term still contributes, and $\gamma_{\mathrm{eff}}^{\mathrm{int}} \approx 0.035$.

While cascading does not provide a means for making the nonlinear-optical response larger than the fundamental limit for a single molecule with the same number of electrons as the molecule pair, cascading introduces an experimentally controllable parameter, $r$, the distance between molecules.  Thus, cascading may provide an additional degree of freedom for using molecules with a sub-optimal direct second hyperpolarizability to make a material with a larger effective second hyperpolarizability.  Indeed, the cascading measurements found in the literature are most likely operating in this domain.

Our results suggest that combining two molecules into a single molecule of optimal structure should yield the largest response. Connecting more molecules together into a super molecule is an even better approach.  We have shown that the fundamental limit of the second hyperpolarizability scales as the seventh power of the length, compared with the model of Rustagi and Ducuing, which shows scaling in proportion to the fifth power - consistent with conjugated carbon chains.  Thus, it should be possible to make materials with much larger nonlinear response.  However, the paradigm for doing so needs to be identified, and as we have seen in the present work, the sum rules and fundamental limits provide a useful guide for assessing the potential for a given approach.

{\bf Acknowledgements}  We thank the National Science Foundation (ECCS-0756936) and the Air Force Office of Scientific Research (FA9550-10-1-0286) for generously supporting this work.

\bibliographystyle{model1a-num-names}

\end{document}